\documentclass[pre,floatfix,showpacs,twocolumn]{revtex4}
\usepackage{color}
\usepackage{epsfig}
\usepackage{graphicx}
\usepackage{amsmath}

\newcommand {\be}  {
\begin{equation}
}
\newcommand {\ee}  {
\end{equation}
}
\newcommand {\bea} {
\begin{eqnarray}
}
\newcommand {\eea} {
\end{eqnarray}
}

\newcommand{\erfc}{\mathrm{erfc}}

\begin{document}

\title{The influence of short range forces on melting along grain boundaries}

\author{C. H\"uter}
\affiliation{Computational Materials Design Department, 
Max-Planck Institut f\"ur Eisenforschung, D-40237 D\"usseldorf, Germany}
\author{F. Twiste}
\affiliation{Computational Materials Design Department, 
Max-Planck Institut f\"ur Eisenforschung, D-40237 D\"usseldorf, Germany}
\author{R. Spatschek}
\affiliation{Computational Materials Design Department, 
Max-Planck Institut f\"ur Eisenforschung, D-40237 D\"usseldorf, Germany}
\author{E. A. Brener}
\affiliation{Peter-Gr\"unberg-Institut 2, 
Forschungszentrum J\"ulich, D-52425 J\"ulich, Germany}
\author{J. Neugebauer}
\affiliation{Computational Materials Design Department, 
Max-Planck Institut f\"ur Eisenforschung, D-40237 D\"usseldorf, Germany}


\date{\today}

\begin{abstract}
We investigate a model which couples diffusional melting and nanoscale structural forces via a combined nano-mesoscale description.
Specifically, we obtain analytic and numerical solutions for melting processes at grain boundaries influenced by structural disjoining forces in the experimentally relevant regime of small deviations from the melting temperature. 
Though spatially limited to the close vicinity of the tip of the propagating melt finger, the influence of the disjoining forces is remarkable and leads to a strong modification of the penetration velocity. 
The problem is represented in terms of a sharp interface model to capture the wide range of relevant length scales, predicting the growth velocity and the length scale describing the pattern, depending on temperature, grain boundary energy, strength and length scale of the exponential
 decay of the disjoining potential.
Close to equilibrium the short-range effects near the triple junctions can be expressed through a contact angle renormalisation in a mesoscale formulation. 
For higher driving forces strong deviations are found, leading to a significantly higher melting velocity than predicted from a purely mesoscopic description.
\end{abstract}

\maketitle

\section{Introduction}

The presence of grain boundaries in a vast range of materials used in metallurgical processes has crucial influence on their key features. Elevated temperatures during the 
processing and reduced local transition temperatures can lead to phenomena of grain boundary induced failure, as e.g.~hot cracking \cite{MRappazAJacotWJBoettinger_MetallMaterTransA_34_467_2003}. 
Grain boundary melting is of importance in many industrial processes nowadays, as materials with specifically low grain boundary melting temperatures become widely used.
The behaviour of superheated grain boundaries was investigated on atomistic scales in \cite{WFanXGong_PRB72_064121_2005}, where symmetric tilts showed an extended regime of stability above the melting temperature.    
Grain boundary premelting was investigated recently e.g.~in \cite{AAdland_etal_PRB87_024110_2013,NWang_etal_PRE81_051601_2010, RSpatschek_etal_PRB87_024109_2013, YMishin_etal_ActaMat57_3771_2009},
showing that for low misorientation an attractive interaction between adjacent solid-melt interfaces can stabilise a grain boundary.
Recent investigations of heterogeneous nucleation of liquid droplets in overheated grain boundaries  link the short range interactions to nucleation processes \cite{TFrolovYMishin_PRL106_155702_2011}.
Up to now, the subsequent melting process along the grain boundary following the nucleation regime has not been investigated, taking into account the influence of the short range forces.
For an overheated crystal, the melt phase becomes wide far behind the triple junction, and a mesoscale perspective has been developed in \cite{EAB_CH_DP_DET_PRL99_105701_2007}.
In contrast to a classical Mullins grooving \cite{Mullins} a steady state growth regime is found here.
This treatment does not resolve the behavior near the tip region which is influenced by microscopic effects in the spirit of  \cite{TFrolovYMishin_PRL106_155702_2011, SnoijerAndreottiPhysFluid20_057101_2008}.
In fact, a combined treatment of the short scale interaction effects and the mesoscale diffusion limited melting process has not yet been achieved and will be the subject of this paper.
A major result will be the establishment of a quantitative link between an effective mesoscale description and the near tip behavior, using the scale-bridging approach developed here, which is valid for low overheating.
As a result, we obtain a closed description, which contains in a single framework the short-ranged interactions on the nanometer scale and simultaneously captures the kinetics of the melting process, which is a typical mesoscale process.
We note that this typically demands to resolve multiple characteristic lengthscales, which typically differ by several orders of magnitude.

Apart from a numerical treatment, which resolves efficiently the phenomena on all these relevant length scales, we also provide an analytical description, which is in excellent agreement with the full model.
Its central benefit is that it delivers a closed expression of the melting velocity, and therefore gives deeper insights into the dependencies on the different control and material parameters.

\section{Model description}

\begin{figure}
\begin{center}
\includegraphics[width=9cm]{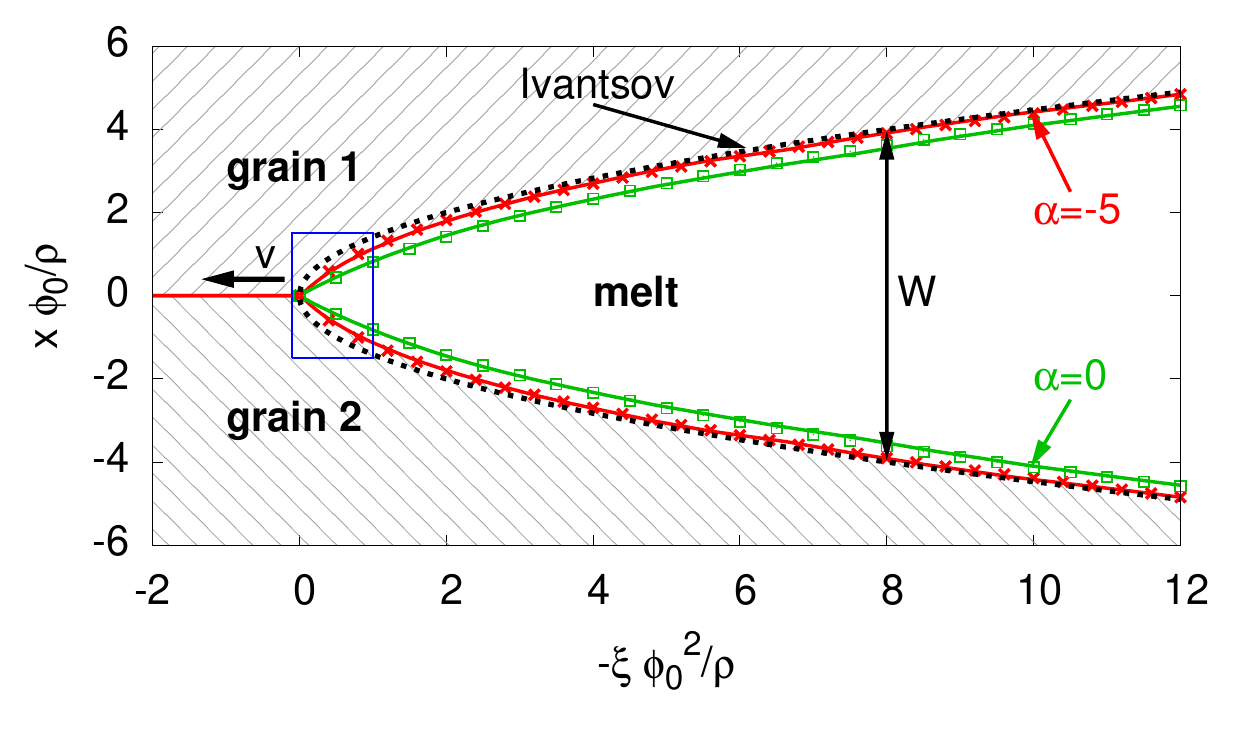}
\includegraphics[width=9cm]{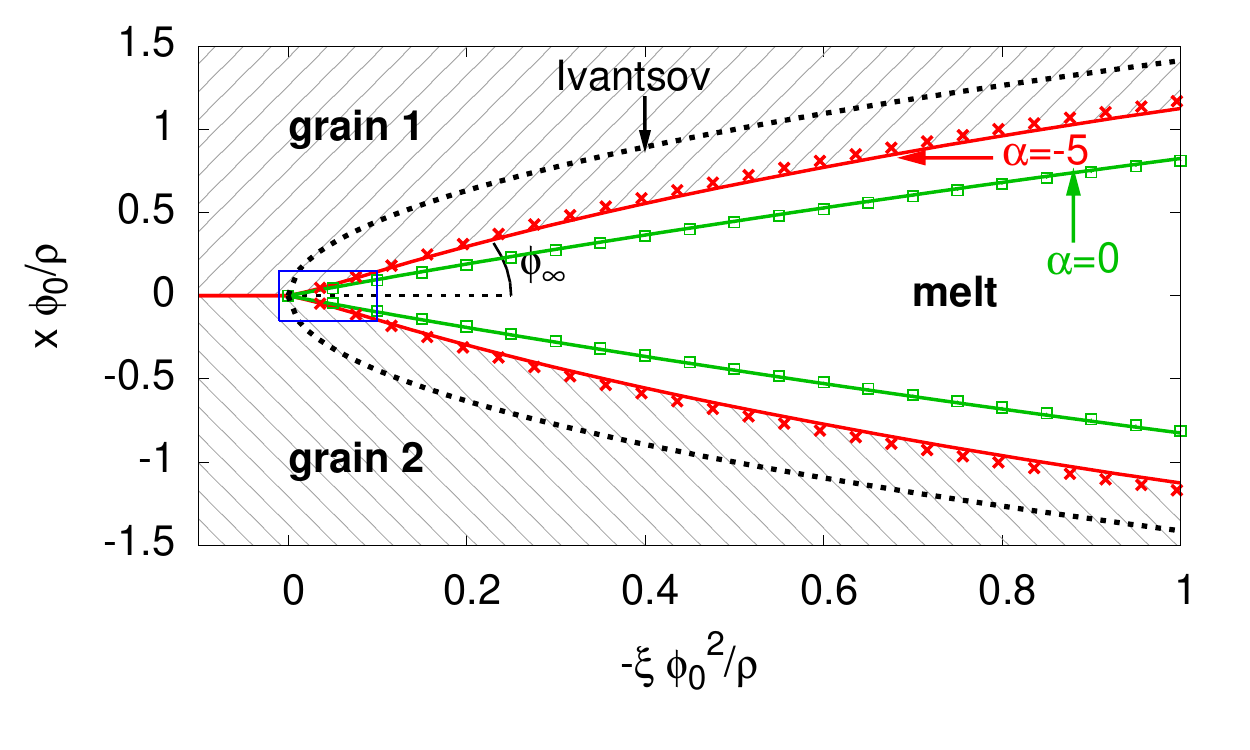}
\includegraphics[width=9cm]{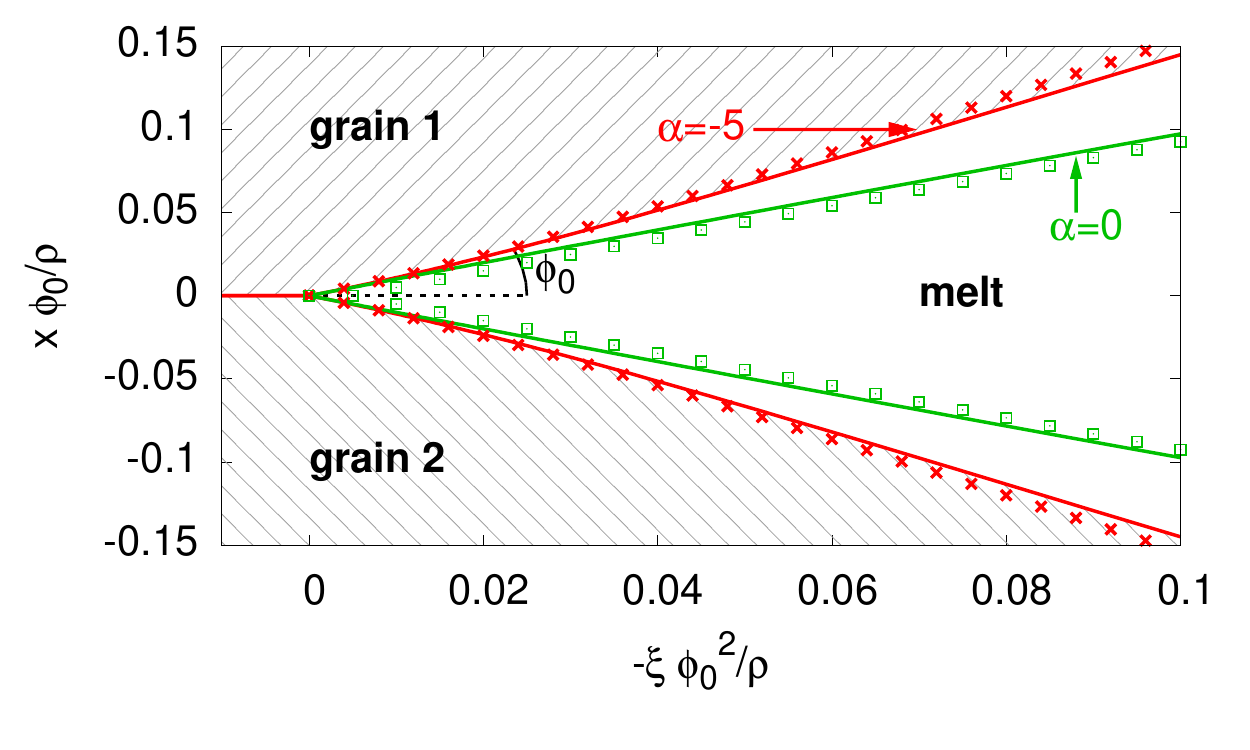}
\caption{\label{geometry}
Melting along an overheated (dry) grain boundary, seen on three different scales.
On the largest scale, the behavior is dominated by the diffusion limited growth (top), on intermediate scales by the (mesoscopic) finite contact angle at the triple junction (centre), and on microscopic scales (bottom) by the solid-melt interface interaction, which bends the interfaces.
Far behind the triple junction the influence of a finite tip angle and the short range interactions has decayed, and the contours approach an Ivantsov parabola (black dotted curve).
Solid curves result from a numerical solution of the full problem (\ref{BIRepresentation}), points from the analytical solution of Eq.~(\ref{linEq}).
We find good agreement between the full numerical solution and the analytical approximation on all scales.
The axes are scaled by the asymptotic parabola radius $\rho$ and additionally by the microscopic tip angles $\phi_0$ and $\phi_0^2$, in agreement with the theoretical approach and better visibility of the thin melt front.
The region inside the blue rectangles is magnified in the following subfigure. 
On the intermediate scale, the mesoscopic contact angle $\phi_\infty$ appears, which differs from the microscopic contact angle $\phi_0$ in the presence of short ranged interface interactions (bottom panel).
In all cases the microscopic tip angle is $\phi_0=0.02$, the overheating $\Delta=0.01$ and the range of the short range interactions $\beta=4$.
}
\end{center}
\end{figure}

We describe the interaction of structural forces on the nanoscale and diffusional grain boundary melting via a sharp interface model in terms of boundary integral equations.
The geometry of the system is shown in Fig.~\ref{geometry}.
Here a melt finger is propagating into the overheated solid along a grain boundary, and we expect that the presence of short range structural interaction strongly influences  the kinetics of the process.
On large scales (top panel), a mesoscopic melt front is advancing along the dry grain boundary, with a characteristic parabolic front profile.
Closer inspection of the tip region (middle panel) shows the appearance of finite contact angles between apparently straight interfaces.
On the microscopic level (bottom panel) short ranged effects lead to curved interface profiles.

The effects on the atomistic scale are incorporated in terms of the disjoining potential $V(W)$, where $W$ is the local width of the liquid as depicted in Fig.~\ref{geometry}.
It expresses the structural short-range interaction between the two misoriented grains which are separated by a melt layer.
A repulsive interaction, $V'(W)<0$, typical for large misorientations, gives raise to grain boundary premelting, whereas attractive interactions, $V'(W)>0$, stabilise a dry grain boundary.
While combinations of these two cases with extrema in the disjoining potential can occur and result from the structure of $V(W)$ as being a superposition of exponentially decaying contributions with different ranges \cite{NWang_etal_PRE81_051601_2010,AAdland_etal_PRB87_024110_2013,RSpatschek_etal_PRB87_024109_2013}, we focus here on the elementary case of monotonic disjoining potentials.

From a mesoscopic perspective the grain boundary premelting is characterised by $\bar{\gamma} = \gamma_{gb} - 2\gamma_{sl}$, where $\gamma_{gb}$ is the energy of a dry grain boundary and $\gamma_{sl}$ the solid-melt interfacial energy.
It describes the preference of having two solid-melt interfaces versus a grain boundary.
Apparently, $\bar{\gamma} > 0$ represents repulsive grain boundaries, and $\bar{\gamma} < 0$ corresponds to attractive grain boundaries. 
We can then write $V(W)= \bar{\gamma} f(W/\delta)$, where $\delta$ is the atomistic length scale characterising the range of the structural forces, and $f(W/\delta) = \exp (-W/\delta)$. 
This type of effective model for the structural disjoining potential as described recently via atomistic studies in \cite{Fensin_2010_PRE_81_031601}, was originally suggested for the understanding of wetting transitions in \cite{LipowskiFisher_PRB36_2126_1987}.
A seminal description for the appearence of wetting transitions in grain boundaries is found in \cite{KikuchiCahn1980}, further discussions of the relation to structural forces in \cite{JMellenthinAKarmaMPlapp_PRB_78_184110_2008,deGennesRMP1985}.

Together with the local melting temperature shift due to the disjoining potential we take the Gibbs-Thomson effect for curved solid-melt interfaces into account, so in local equilibrium the transition temperature $T_I$ of a melting front is given by
\bea \label{physicalLocEq}
T_{I} = T_M\Bigg[1 + \frac{\gamma_{sl} \kappa}{L} + \frac{\bar{\gamma}}{L \delta} f'(W/\delta) \Bigg],
\eea
where $\kappa$ is the curvature (positive for a convex liquid phase), $L$ the latent heat and the last term the shift of the melting temperature $T_M$ due to the structural forces (see Appendix \ref{LEQ} for a derivation). 
We rephrase the problem in dimensionless units and obtain for the temperature at the solid-melt interfaces
\bea \label{dimLessLocEq}
u|_{int} = \Delta - \Delta_w f' - d\kappa.
\eea
Here, we introduce $u = (T_\infty - T)c_p/L$, $\Delta = (T_{\infty}-T_M)c_p/L$, $\Delta_w = T_M \bar{\gamma} c_p/(L^2\delta)$ and the capillary length $d = T_M \gamma_{sl} c_p/L^2$, using the heat capacity $c_p$. 
$T_\infty$ is the temperature which is applied far away from the grain boundary.
For melting processes, we have $T_\infty>T_M$, and therefore far behind the triple junction, where also the short-range interactions have decayed, a parabolic profile is found \cite{EAB_CH_DP_DET_PRL99_105701_2007}, as shown in the top panel of Fig.~\ref{geometry}.

Heat transport is described by the bulk diffusion equation 
\be
\label{govEqs1}
D \nabla^2 u = \partial u/\partial t,
\ee
with the diffusion constant $D$ which is assumed to be equal in the solid and liquid phase. 
The difference between the temperature gradients on the solid (S) and liquid (L) side of the interface accounts for the latent heat consumption at a propagating interface,
\be
\label{govEqs2}
v_n = D \vec{n}\cdot \left(  \nabla u_L - \nabla u_S \right)|_{int},
\ee
with the interface normal $\vec{n}$ and the normal component $v_n$ of the interface velocity.

The equivalent Green's function formulation of the moving boundary problem is more convenient for our purposes in view of a combined numerical and analytical treatment \cite{JSLangerLATurski_ActaMetall_25_1113_1977, CHueter_GBoussinot_EABrener_PRE_83_050601_2011, TFischaleckKKassner_EPL_81_54004_2008}, as well as by the need to resolve the dynamics of the process on various scales as depicted in Fig.~\ref{geometry}. 
Thus, eliminating the thermal field in Eq.~(\ref{dimLessLocEq}) and rescaling all lengths by the tip radius of curvature $\rho$ of the asymptotic parabola, we obtain in a comoving frame of reference for a steady state solution
\bea
\label{BIRepresentation}
\Delta + \Delta_w \exp\left(\frac{- 2 |{x}| \rho}{\delta} \right)- \frac{d}{\rho}  \kappa = \\ \nonumber
 \frac{p}{\pi} \int_{-\infty}^{\infty} dx' e^{-p(\xi - \xi')} K_0 \left( p |\vec{r}-\vec{r}'|  \right)
\eea 
with the integration along the solid-melt interface, parametrized by the dimensionless function $\xi(x)$, see Fig.~\ref{geometry}.
This formulation combines Eqs.~(\ref{dimLessLocEq})-(\ref{govEqs2}) in closed form as a nonlinear eigenvalue problem to determine the interface contour and the scale $d/\rho$. 
By $K_0$ we denote the modified Bessel function of third kind in zeroth order. 
The Peclet number $p = \rho v/(2 D)$ is the ratio of the tip radius of curvature of the asymptotically matched parabola in the region $\xi\to-\infty$ and the diffusion length $2 D/v$.
At the tip, the description is supplemented by the knowledge of the contact angle, and this will be discussed in more detail below.
The tail region, where the influence of the triple junction and the short-range interaction is no longer relevant, determines the relation between the overheating $\Delta$ and the Peclet number via the classical Ivansov relation $\Delta = \sqrt{\pi p}\, e^p \mathrm{erfc}(\sqrt{p})$ \cite{Ivantsov}.
This is an important ingredient, as this relation, which appears on the largest scale of the model, controls the growth velocity depending on the parabola radius $\rho$.
In turn, the short ranged effects, which can affect the growth kinetics, therefore also influence the asymptotic front profile, hence coupling short-range effects with the mesoscopic front profiles.

\section{Contact angle renormalisation}
\label{reno::section}

We begin our analysis of the model by consideration of an equilibrium situation, $\Delta=0$, for an attractive grain boundary, i.e.~$\bar{\gamma}=V(0)<0$, where the interfaces are stationary.
Then, in Eq.~(\ref{BIRepresentation}) also the integral term, which expresses the latent heat absorption at solidifying fronts, vanishes.
On scales, which are large in comparison to the microscopic scale $\delta$, the short range interactions have decayed, and straight interfaces form, in order to minimise the interfacial energy.
Seen on this larger scale, these straight interfaces come together at the triple junction, where they form a {\em mesoscopic} contact angle $\phi_\infty$, as shown in the second panel of Fig.~\ref{geometry}, which in full equilibrium is given by Young's law (we consider only isotropic surface energy, and therefore torque terms do not show up).
It reads in small angle approximation, $|x'|\ll 1$,
\begin{equation} \label{young}
\bar{\gamma}= -\phi_\infty^2\gamma_{sl}.
\end{equation}
This relation anticipates that the triple junction is mobile, and the total interfacial free energy energy is minimised with respect to this degree of freedom.

On scales $W\sim \delta$ the short-ranged interface interaction sets in and bends the solid-melt interfaces.
We expect for attractive interactions convex solid phases, as this effectively brings the solid-melt interfaces closer to each other and therefore reduces the energy.
As a result, the mesoscopic contact angle $\phi_\infty$ on scales $W\gg \delta$ deviates from the {\em microscopic} angle $\phi_0$, which is defined at the very tip as $\tan\phi_0 = - x'(\xi=0)$, as illustrated in the bottom panel of Fig.~\ref{geometry}.

For $\Delta=0$ the condition (\ref{BIRepresentation}) can also be interpreted as minimisation $\delta F/\delta x(\xi)=0$ of the energy functional
\begin{equation}
F = \int^{0}_{-\infty} \left[ 2\gamma_{sl}  \left( 1+ x'^2 \right)^{1/2} + V(W)\right] d\xi,
\end{equation}
which consists of the energy of the two solid-melt interfaces and the disjoining potential $V(W)$.
The prime denotes differentiation with respect to $\xi$.
The origin $x=\xi=0$ is chosen as the position of the triple junction.
In small slope approximation for narrow melt fingers the equilibrium condition becomes $\gamma_{sl}[x'(\xi)^2]'=[V(2x(\xi))]'$, in agreement with Eq.~(\ref{BIRepresentation}).
Integration yields
\begin{equation} \label{reno}
\phi_\infty = \sqrt{\phi_0^2 - \frac{\bar{\gamma}}{\gamma_{sl}}}.
\end{equation}
Hence for an attractive interaction, $\bar{\gamma}<0$, the mesoscopic contact angle $\phi_\infty$ is larger than the microscopic one, $\phi_0$, in agreement with our expectation.
From the comparison of Eqs.~(\ref{young}) and (\ref{reno}) we obtain in full equilibrium $\phi_0=0$.
This result is a natural consequence of the continuous interpolation of the solid-melt interface energy $2\gamma_{sl}$ to the grain boundary energy $\gamma_{gb}$ if the two interfaces come closer.
Therefore, also the interface curvature changes continuously at the triple junction from the wet side to the dry grain boundary, i.e.~$\phi_0=0$.
However, as mentioned before, this anticipates full equilibration of the triple junction, and Eq.~(\ref{reno}) can be understood as a generalisation of Young's law (\ref{young}).

In general, the behaviour of the triple point will be controlled by independent kinetics, which are not in the focus of the present work.
This implies, that for growth situations, the triple junction may not fully equilibrate, which can lead to finite tip angles $\phi_0$, as discussed also in Appendix \ref{TrijunctionKinetics}.
For a more thorough discussion of this issue we refer to \cite{SnoijerAndreottiPhysFluid20_057101_2008}.
As a generalisation, we therefore allow also for microscopic angles $\phi_0>0$ in the discussion of melting in the following section.
Moreover, we mention in passing that in general also other effects can lead to modified contact angles, among them surface roughness and heterogeneities, see e.g.~\cite{AMarmur_AdvCollIntSci50_121, RNWenzel_IndEngChem28_p988_1936,GWolonskyAMarmurCollSurfA_156_p381_1999,LGaoTJMcCarthy_Langmuir2007_23_3762}.

\section{Results and Discussion of the entire melting process}
We split our approach threefold when we focus on the melting process--- in the most general regime we solve the boundary integral formulation of the problem as stated in Eq.~(\ref{BIRepresentation}) numerically, depending on the overheating for various disjoining potential parameters. As pointed out in detail later in this section, we then distinguish two limiting regimes, where we can simplify the governing equations and reduce the problem such that we can predict both the eigenvalue and the interface analytically.
Finally, we relate the results obtained numerically by the direct solution of Eq.~(\ref{BIRepresentation}) to the results we obtained analytically and semi-analytically in the limiting regimes of the parameter space accessible at small overheatings. 

We begin with the results from the direct numerical solution of Eq.~(\ref{BIRepresentation}). In this representation, the problem demands the solution of the interface shape $\xi(x)$ such that it matches the prescribed slope at the tip and the parabolic asymptotics far behind the triple junction. 
The obtained eigenvalue $d/\rho$ as function of the overheating $\Delta$, the magnitude of the disjoining potential $\Delta_w$, the ratio of the lengthscale of the disjoining potential and the capillary length, $\delta/d$, and the opening angle $\phi_0$ allows to extract the melting velocity as $v = 2D p(\Delta)/\rho$. 
For the sake of clarity, we show the obtained results for Eq.~(\ref{BIRepresentation}) in terms of $\mu = d \phi_0^3/(\rho\Delta)$, which effectively expresses the growth velocity $v$ as function of $p/\phi_0^2$ as measure for the overheating as driving force for different values of $\alpha = \Delta_w/\Delta$ (strength of interaction) for fixed $\beta = 2 d \phi_0^2/(\delta \Delta)$ (range of interaction). 
This is the reduced set of independent parameters which suffices to describe the problem in the analytic calculations below, and we can thus easily compare the outcome of all approaches. 

To estimate the range of the appearing parameters, we start with typical interfacial energies $\bar{\gamma}\sim\gamma_{sl}\sim\gamma_{gb}\sim 10^{-2}\,\mathrm{J/m^2}$.
The range of the disjoining forces is typically in the range of $\delta\sim10^{-10}\,\mathrm{m}-10^{-9}\,\mathrm{m}$.
Using values for aluminium, $T_M=660\,\mathrm{K}$, $L\approx8.7\cdot10^8\,\mathrm{J/m^3}$ and an overheating $(T_\infty-T_M)/T_M\sim 0.01-0.1$ gives $\alpha\sim 1-10$.
With $d\sim10^{-9}\,\mathrm{m}$ we have $\beta\sim 1$.

For the example of $\delta$ iron near the melting point we can use more explicitly the disjoining potential determined from amplitude equations descriptions for a symmetric tilt [100] grain boundary with a misorientation of $11.4^\circ$.
The functional form $V(W)=\bar{\gamma}\exp(-W/\delta)$ fits well to the attractive tail of the interaction in Ref.~\onlinecite{AAdland_etal_PRB87_024110_2013} with $\bar{\gamma}=-610\,\mathrm{mJ/m^2}$ and $\delta=0.18\,\mathrm{nm}$ and solid-melt interface energy $\gamma_{sl}= 144\,\mathrm{mJ/m^2}$ \cite{Spatschek2010}.
These estimates demonstrate the applicability of the description to a wide class of metallic systems.

\begin{figure}
\begin{center}
\includegraphics[width=8.5cm]{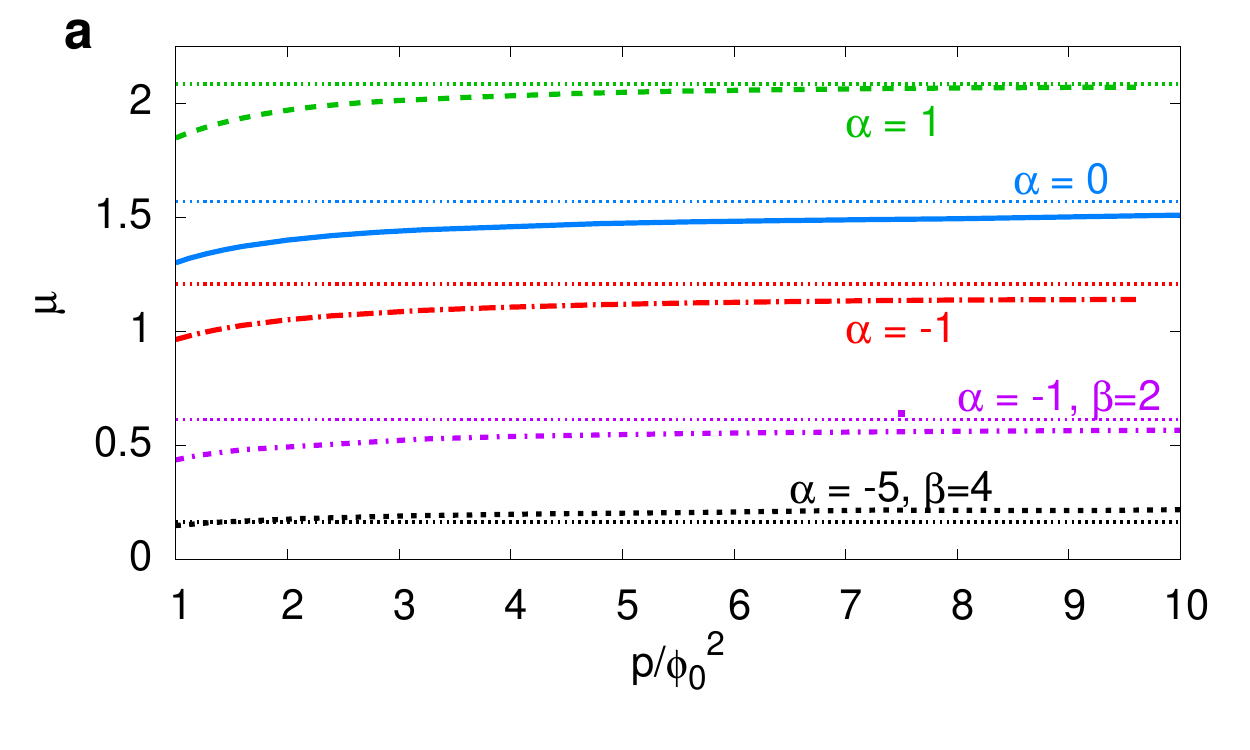}
\includegraphics[width=8.5cm]{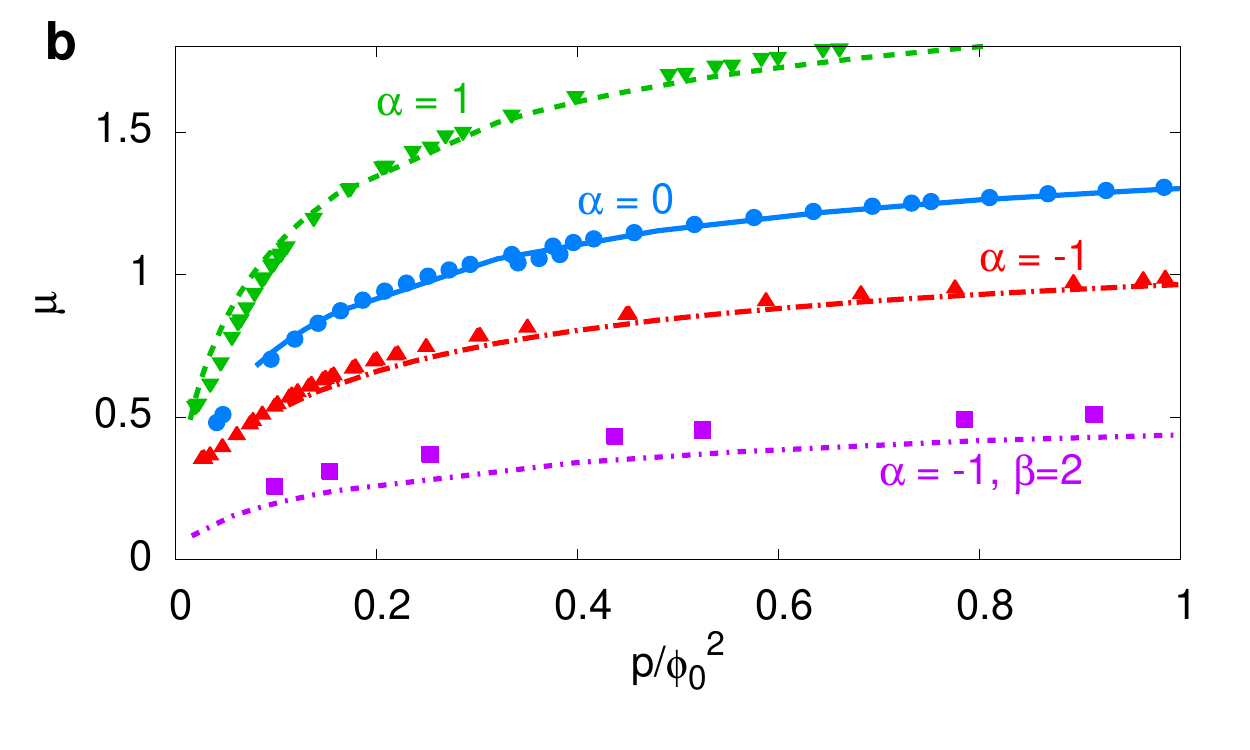}
\caption{\label{fullMu} 
The eigenvalue $\mu = d \phi_0^3/(\rho\Delta)$ as function of $p/\phi_0^2$ for $\beta = 10$ (if not stated differently) and $\phi_0=0.025$. 
A larger eigenvalue corresponds to a higher melting velocity.
The curves result from the solution of the full problem (\ref{BIRepresentation}), the symbols from the linearised description (\ref{linNumRepresentation}), and the data matches for small $p/\phi_0^2$ (in panel b).
For higher values of $p/\phi_0^2$ (panel a), the results converge towards the the analytical solutions of Eq.~(\ref{mu_cubeRoot}), which are shown as dotted horizontal lines in panel a. 
}
\end{center}
\end{figure}

The resulting plot is split into two regimes, see the continuous curves in Fig.~\ref{fullMu}, for large ratios of Peclet number and opening angle $p/\phi_0^2$ in the top part, and for small values in the bottom part.  
We find that attractive interactions $\alpha<0$ lead to a smaller eigenvalue $\mu$ and growth velocity, in agreement with the intuitive expectation.

Beyond the purely numerical approach for Eq.~(\ref{BIRepresentation}), we focus on narrow melt fronts with small opening angles $\phi_0\ll 1$ and small overheating, $\Delta\ll 1$. 
As seen in Fig.~\ref{geometry}, $-d\xi/dx = (\tan \phi_0)^{-1}$, and we rescale such that $-d\xi/dx = 1$, i.e.~$x \to x\rho/\phi_0$, $\xi \to \xi\rho/\phi_0^2$, with Ivantsov asymptotics, $\xi_{Iv} \simeq - x^2/2$.
In conjunction with the exchange of the dependent and independent variables, $x \leftrightarrow \xi$, this allows to linearize the curvature term and the integral kernel in Eq.~(\ref{BIRepresentation}).
For the exponential contribution from the disjoining potential, the argument then reads $-\beta \xi/\mu$, which decays within the close vicinity of the triple junction, such that we can approximate $-d\xi/dx \approx 1$ there.
We note that due to the renormalisation of the contact angle the interface is curved on the scale $\delta$ near the origin, and therefore the assumption of a straight behaviour for the exponential term is only approximative.
This allows us to obtain an equation which is entirely linear in $dx/d\xi$,
\bea
&& 1 + \alpha \exp\left( -\frac{\xi \beta}{\mu}\right) + \mu \frac{d^2 x}{d\xi^2} \label{linNumRepresentation} \\
&&= \frac{2p^{1/2}}{\pi^{3/2} \phi_0}   \int\limits_0^{\infty} d\xi' \frac{dx}{d\xi'} \exp\left(-\frac{p(\xi' - \xi)}{\phi_0^2}\right) K_0 \left( \frac{p |\xi' - \xi|}{\phi_0^2} \right). \nonumber   
\eea
We solve Eq.~(\ref{linNumRepresentation}) numerically for the slope of the interface profile.
The resulting eigenvalue $\mu$ as function of $p/\phi_0^2$ is shown as isolated points in Fig.~2, exhibiting an excellent agreement with the solution of the full problem Eq.~(\ref{BIRepresentation}).

Close to the origin, $p/\phi_0^2\ll 1$, we expect for the eigenvalue $\mu$ the scaling $\mu \sim \sqrt{p/\phi_0^2}$ without short ranged interactions ($\alpha=0$) due to the structure of Eq.~(\ref{linNumRepresentation}), see \cite{EAB_CH_DP_DET_PRL99_105701_2007}.
This equation
suggests that the influence of the short range forces decays on the scale $\mu/\beta$, so the relative range of the structural forces is exponentially small in the limit of small $p/\phi_0^2$.
Consequently, we assume that the dependence of $\mu$ on $\alpha$ reduces to exponentially small corrections and preserves the above scaling $\mu \sim \sqrt{p/\phi_0^2}$ in this regime.
Based on this we can find also $\sigma := d/(\rho p)\sim 1/\phi_0^4$ (for $\phi_0\ll 1$).
This is in general agreement with the theory of dendritic growth which states that for finite value of $\sigma$ a cusp ($\phi_0<\pi/2$) appears at the origin of the melt front for isotropic surface tension \cite{EfimAIP}.

When we consider the specific case $p/\phi_0^2 \gg 1$ within the regime $\Delta\ll 1$, $\phi_0\ll 1$, the asymptotic approximation for large arguments of the modified Bessel function holds, and we can simplify the integral kernel in Eq.~(\ref{linNumRepresentation}) and also cut the range of integration to the point of observation.
The obtained Volterra integro-differential equation, 
\bea \label{linEq}
1 + \mu \frac{d^2x}{d\xi^2} + \alpha \exp(-\frac{\beta}{\mu}\xi)= \frac{\sqrt 2}{\pi} \int_{0}^{\xi} \frac{1}{(\xi - \xi')^{1/2}} \frac{dx}{d\xi'},
\eea
is accessible via Laplace transform techniques (for details see Appendix \ref{eigenval::appendix}), and the eigenvalue $\mu$ is determined as 
\bea \label{mu_cubeRoot}
\mu &=&  \frac{\pi}{2}  \left[ \frac{1+\alpha-\beta}{2}  + \sqrt{\left(\frac{1+\alpha-\beta}{2} \right)^2 + \beta} \right]^3.
\eea
The selection appears here through the necessity to suppress exponentially growing modes in the tail of the melt front profile, as discussed in \cite{EAB_CH_DP_DET_PRL99_105701_2007}.

We compare for three different ratios $\alpha = \Delta_w/\Delta$ the solution obtained from Eq.~(\ref{mu_cubeRoot}) to the values for $\mu$ which we obtained via the direct numerical solution of Eq.~(\ref{BIRepresentation}) for $p/\phi_0^2 \gg 1$ in Fig.~\ref{fullMu}.
Here the top panel shows how the analytically predicted eigenvalues are asymptotically reached by the numerically calculated eigenvalues of the full problem (\ref{BIRepresentation}). 
Since $\mu\simeq d\phi_0^3 v\pi/(2D\Delta^3)$ for low overheating ($\Delta\simeq (\pi p)^{1/2}$ there), the velocity is proportional to the eigenvalue $\mu$, and allows to easily convert back the dimensionless parameters to observable quantities, see Appendix \ref{eigenval::appendix}.

Overall, we find excellent agreement of the eigenvalues predicted by all three approaches to the problem.

Finally, we also calculate the interface contour analytically in the regime  $\phi_0^2 \ll 1$, $\Delta^2 \ll 1$, $p/\phi_0^2 \gg 1$. 
For this purpose we solve the inverse Laplace transformation problem for the slope of the interface, as described in Appendix \ref{contour::app}.
The integrated slopes yields the interface profiles shown in Fig.~\ref{geometry} with attractive and without short range interactions.
First, for the given parameters we find an excellent agreement with the unapproximated numerical solution in the entire regime.
Second, increasing the magnitude of the structural forces leads to increasingly wider tails for attractive interactions.
This is a result of the long-range transport, which fixes the product $\rho v/2D$ via the Ivantsov relation.
Hence a slower front demands a wider tail.

\section{Connection between microscopic and mesoscopic descriptions}

\begin{figure}
\begin{center}
\includegraphics[width=9cm]{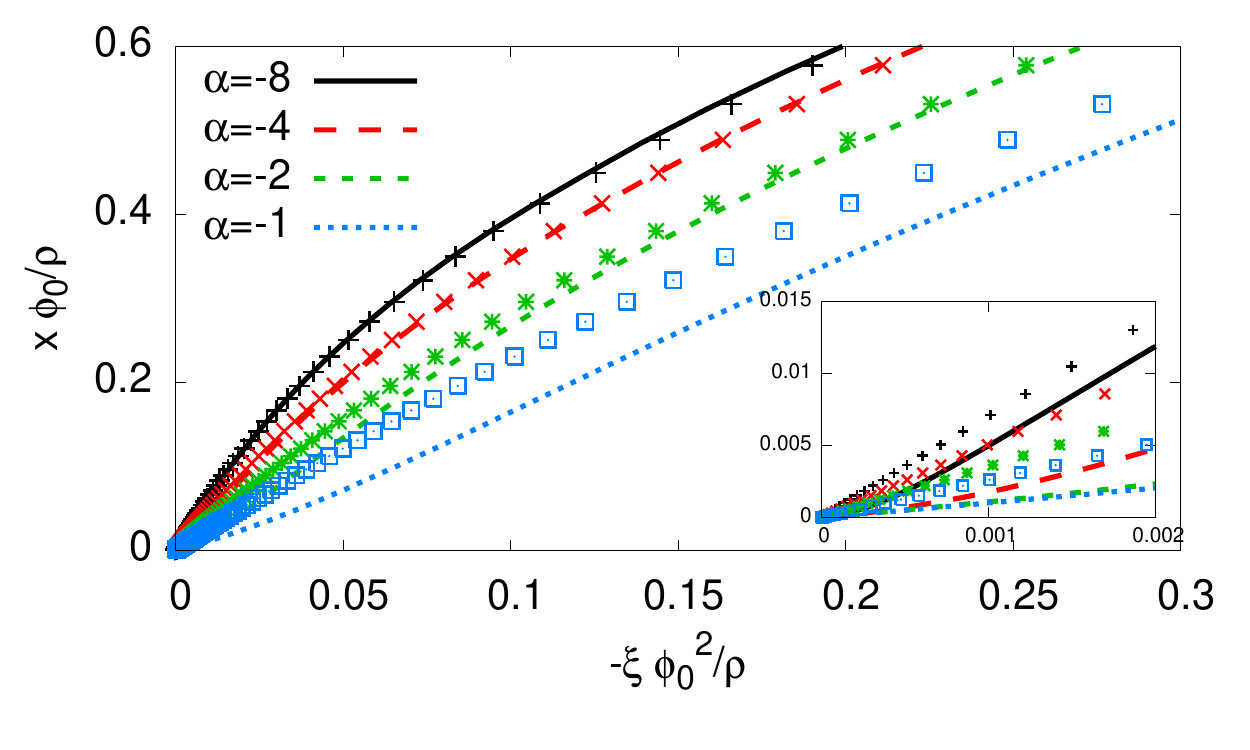}
\caption{
Upper half of the solid-melt front. The plot shows interface contours based on the full model (\ref{BIRepresentation}) including the short-range interaction (continuous curves) and mesoscopic simulations without short-range interactions (shown as symbols), but using the contact angle renormalisation (\ref{reno}).
We use $\Delta=0.01$, $\phi_0=0.01$ and $d/\delta=16.7$, hence $\beta=0.32$.
For large absolute values of $\alpha$ the ratio $\rho/(\phi_\infty\delta)$ is largest, and then the microscopic model (with  tip angle $\phi_0$) delivers the same interface contour as the mesoscopic model (with $\phi_\infty$) on large scales.
Close to the trijunction, the shapes differ, as shown in the inset.
}
\label{fig5}
\end{center}
\end{figure}

In contrast to \cite{EAB_CH_DP_DET_PRL99_105701_2007}, where a purely mesoscopic description of the melting along a grain boundary has been achieved, we consider here additionally the influence of short range interactions, which therefore demand a treatment on a wide range of length scales.
Under certain circumstances, it is possible to directly transfer the microscopic behavior in the tip region to the mesoscopic description \cite{EAB_CH_DP_DET_PRL99_105701_2007}, and this link is analysed here.
It is based on the contact angle renormalisation, as studied in section \ref{reno::section}.
Provided that the melting process is slow, such that the interfaces can almost fully establish equilibrium on short scales $W\sim \delta$, and additionally under the condition that the length scale $\delta$ of the short ranged forces is significantly smaller than the length scale of the parabolic melting front, $\phi_\infty\delta\ll \rho$, the renormalised contact angle appears as effective boundary condition in the mesoscopic description without short ranged forces ($\alpha=0$).
We therefore expect, that under these circumstances, both the mesoscopic description with the {\em effective} boundary condition $x'(0)=-\tan \phi_\infty$ (using $\alpha=0$) and the microscopic model with $x'(0)=-\tan \phi_0$ (with $\alpha<0$), which covers all length scales, should lead to the same results on scales larger than $\delta$.
For that purpose, we have performed two sets of simulations using the full nonlinear model (\ref{BIRepresentation}), and the results are shown in Fig.~\ref{fig5}.
For the values used there, the scale separation varies between $\phi_\infty\delta/\rho=0.03$ for $\alpha=-8$ to $\phi_\infty\delta/\rho=0.53$ for $\alpha=-1$.
The separation of scales is therefore better for larger absolute values of $\alpha$, and then the interface contours obtained from the two approaches coincide on mesoscopic scales;
on short scales of the order $\delta$, there are always deviations, as shown in the inset of Fig.~\ref{fig5}, since the mesoscopic approach contains the short range influence only in an effective sense.
We also extract the eigenvalue $d/\rho$, which is proportional to the melting velocity, as obtained from the two complementary approaches, and find very good agreement also in cases, where the interface contours deviate significantly. 
We find that the obtained eigenvalue $d/\rho=2Dv p/d$ scales in this regime as $d/\rho\sim \alpha^{-2}$.
This scaling is consistent with the mesoscopic prediction $d/\rho \sim \Delta^2/\phi_{\infty}^4 \sim \alpha^{-2}$  which is valid for $\Delta/(\pi\phi_{\infty}) \ll1$, see the discussion above following the linearised description Eq.~(\ref{linNumRepresentation}).

\begin{figure}
\begin{center}
\includegraphics[width=8cm]{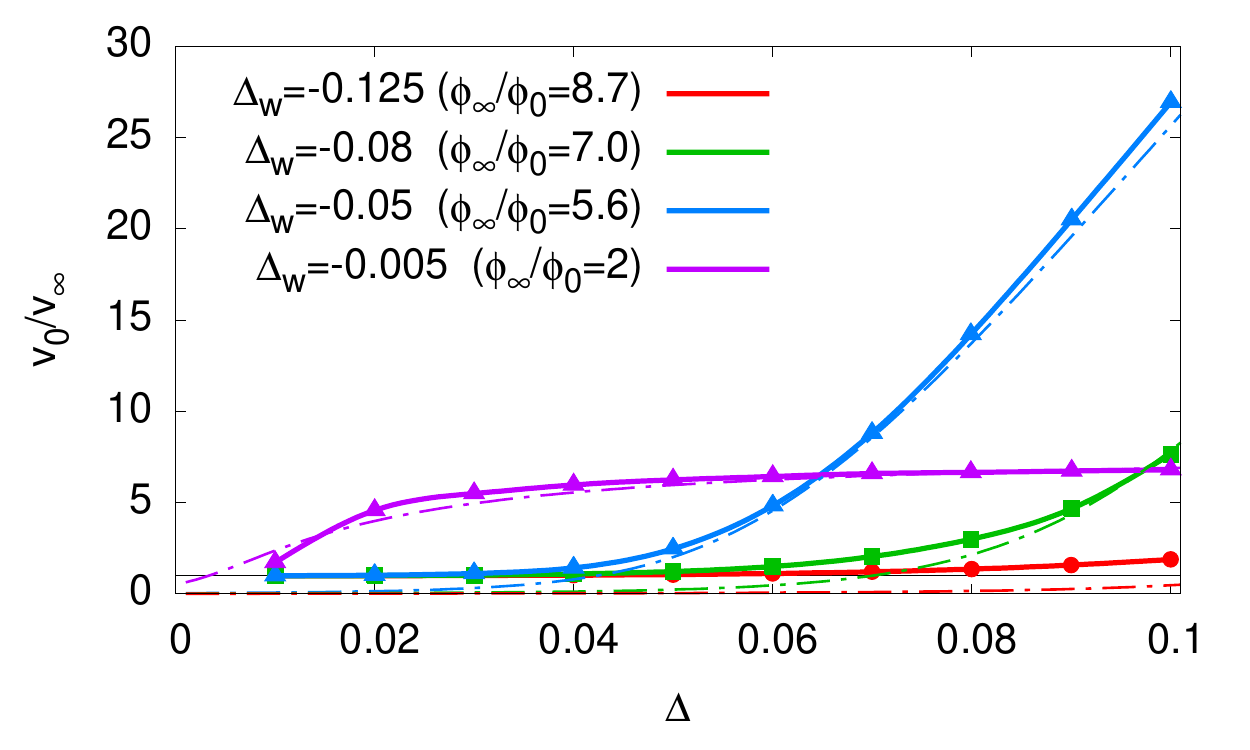}
\includegraphics[width=8cm]{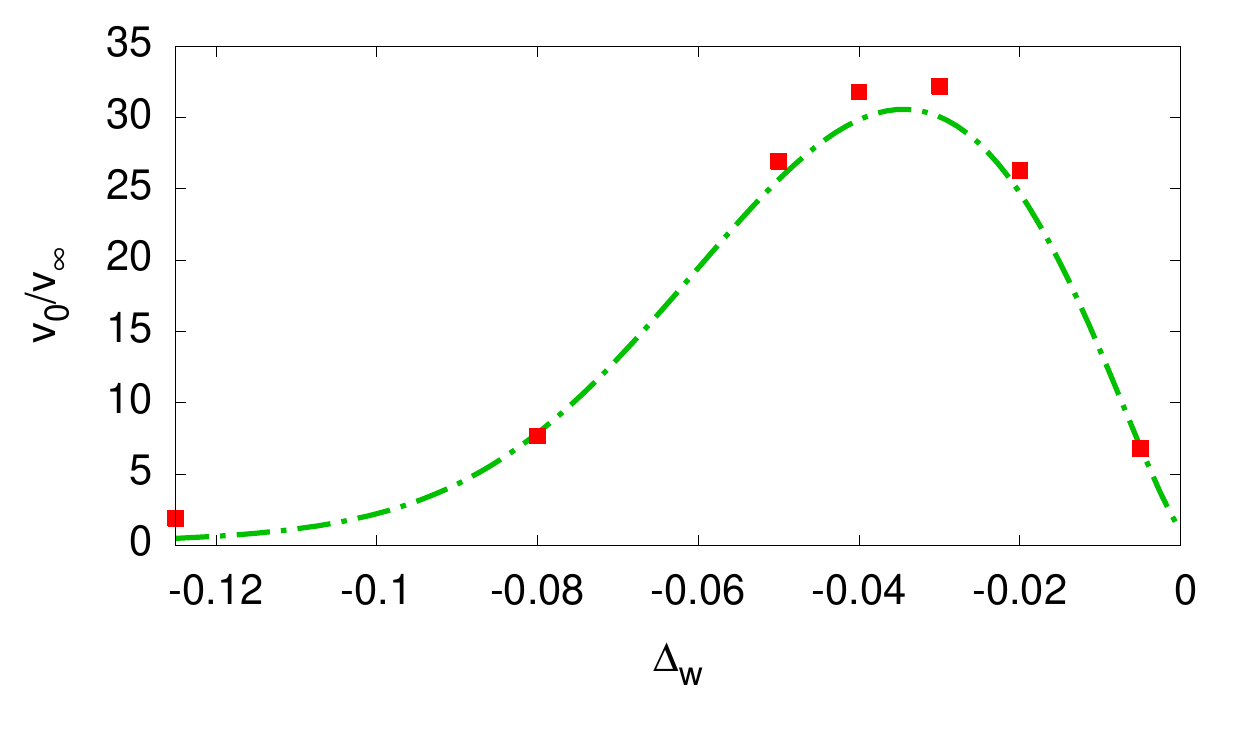}
\caption{Top: Ratio of melting velocities of the microscopic model, $v_0$, and the corresponding velocity for the mesoscopic model, $v_\infty$.
The thick lines and symbols result from the solution of the full nonlinear model (\ref{BIRepresentation}), the thin dashed-dotted line from the analytical expression (\ref{mu_cubeRoot}), showing a good agreement for higher overheatings.
Bottom: Ratio of melting velocity as function of $\Delta_w$ for $\Delta=0.1$.
The red points are results from the full problem (\ref{BIRepresentation}), the dashed dotted curve from the analytical result (\ref{mu_cubeRoot}).
For $\Delta_w=0$ both descriptions coincide, but near the maximum the microscopic description leads to a significantly higher melting velocity.
The remaining parameters are in both panels $\phi_0=0.01$, $d/\delta=16.67$.
}
\label{fig7}
\end{center}
\end{figure}

We point out that the reduction to the mesoscopic model via the contact angle renormalisation implies a strong reduction of the parameter space.
A priori, the melting velocity is a function of four parameters, $v(\Delta, \phi_0, \Delta_w, d/\delta)$, whereas the matching allows to reduce it to only two parameters, $v(\Delta, \phi_\infty)$.
In this regime, the microscopic details are therefore fully contained in the information of the mesoscopic contact angle $\phi_\infty=(\phi_0^2 - \Delta_w \delta/d)^{1/2}$.
To understand better the applicability of this mapping, we performed simulations of the full model (\ref{BIRepresentation}) and corresponding simulations with renormalized contact angle and without short-range interactions.
The ratio of the obtained velocity $v_0/v_\infty$ of the microscopic to the mesoscopic model is shown as function of overheating in Fig.~\ref{fig7}.
Close to equilibrium, both approaches indeed give the same velocity.
For higher driving forces however substantial deviations emerge.
The melting velocity of the full microscopic model then predicts a significantly higher melting velocity than the mesoscopic model.
This demonstrates the striking influence of the short range forces also on the kinetics of grain boundary melting beyond a purely mesoscopic perspective.
The appearance of the maximum in the velocity ratio, as shown in the bottom panel of Fig.~\ref{fig7} is due to two competing effects:
On the one hand, for $\Delta_w/\Delta\to 0$ the short range forces vanish, and therefore the microscopic and the mesoscopic model become the same ($\phi_\infty=\phi_0$).
On the other hand, for low ratios of these driving forces the solid-melt interfaces cannot fully establish locally the equilibrium contact angle renormalisation, and therefore effectively a lower kinetic contact angle $\phi_\infty$ emerges.
It leads to faster growth in agreement with the scaling $v\sim \mu_\infty/\phi_\infty^3$ for fixed driving force $\Delta$.
In turn, for larger strength of the attractive force, i.e.~larger value $-\Delta_w$, hence $\phi_\infty\gg\phi_0$, the microscopic and corresponding mesoscopic model therefore lead again to similar velocities.

The numerical results are also in very good agreement with the closed analytical expression (\ref{mu_cubeRoot}).
For the mesoscopic model ($\alpha=0$) it recovers $\mu_\infty=\pi/2$ (see \cite{EAB_CH_DP_DET_PRL99_105701_2007}), and the velocity ratio is given by
\begin{equation}
\frac{v_0}{v_\infty}=\frac{\mu}{\mu_\infty} \left( \frac{\phi_\infty}{\phi_0} \right)^3. 
\end{equation}
Fig.~\ref{fig7} shows excellent agreement with the full numerical solution and again demonstrates the strong influence of the short ranged forces beyond an equilibrium contact angle renormalisation for larger driving forces.

\section{Summary and Conclusions}

In summary, we have developed a sharp-interface description for steady state melting along a grain boundary, which takes into account the effect of interface interactions near the triple junction. 
This description demands to resolve the process on several orders of magnitude in length.
The reason is that one the one hand the behaviour near the triple junction, where the short ranged interactions are strongest, clearly affects the melting process and the shape of the solid-melt interfaces.
On the other hand, one still has to consider mesoscopic (diffusive) transport on larger scales to predict the kinetics of the process.
Such a scale bridging description is difficult to achieve with other approaches like phase field, as the numerical cost for such simulations would be very high.
Apart from that, our investigations also allow to reduce the complex description to a fully analytical expression, which predicts both the growth velocity and the interface profiles with high accuracy on all scales.
Near equilibrium, the influence of the microscopic short range forces reduces to a renormalisation of the mesoscopic contact angle, which allows to predict the melting velocity and interface profiles quantitatively on distances beyond the scale $\delta$.
For higher overheating, however, the present more detailed microscopic model predicts significantly higher melting velocities than one would expect from a purely mesoscopic consideration.

In conclusion, our findings explicitly show the significant influence of structural nanoscale effects on length and timescales relevant to metallurgical melting processes. 
Hereby, the generic formulation which yields robust scaling laws suggests the importance for whole classes of materials.

\acknowledgments
We acknowledge the support of the Deutsche Forschungsgemeinschaft under Project No. SFB 761.

\appendix

\section{Local equilibrium condition} 
\label{LEQ}

Here we derive the expression for the solid-melt interface temperature given by Eq.~(\ref{physicalLocEq}).
We focus on the disjoining potential here and therefore consider only straight interfaces;
the Gibbs-Thomson term involving the interface curvature can be treated in the usual way.
For a solid-melt-solid layer system at constant temperature $T$ the free energy per unit area is
\begin{equation}
F(T, W) = -L \frac{T-T_M}{T_M} W + V(W)
\end{equation}
with $W$ being the width of the sandwiched melt layer and the disjoining potential $V(W)$ for the interaction between the two solid-melt interfaces.
Minimization of the free energy with respect to $W$ therefore gives
\begin{equation}
T = T_M \left[ 1+ \frac{V'(W)}{L}\right],
\end{equation}
which is the desired expression for $V(W) = \bar{\gamma} f(W/\delta)$.

\section{Trijunction kinetics} 
\label{TrijunctionKinetics}

Young's law expresses full equilibration and therefore minimization of the free energy also with respect to the triple junction position.
In general, it reads as a mesoscopic condition
\begin{equation} \label{young1}
\gamma_{gb} = 2\gamma_{sl}\cos\phi_\infty,
\end{equation}
for $\bar{\gamma}=\gamma_{gb}-2\gamma_{sl}<0$.
It reduces to Eq.~(\ref{young}) for $\phi_\infty\ll 1$.

For finite melting velocity $v$ the equilibrium condition generalises to
\begin{equation} \label{young2}
\gamma_{gb} - 2\gamma_{sl} \cos\phi_\infty = \gamma_{sl} \frac{v}{v_0}
\end{equation}
with a characteristic velocity scale $v_0$.
For $v\ll |v_0|$ the expression recovers Young's law (\ref{young1}).
The solution of Eq.~(\ref{young2}) for $\phi_\infty\ll 1$ becomes
\begin{equation}
\phi_\infty^2 = - \frac{\bar{\gamma}}{\gamma_{sl}} + \frac{v}{v_0}.
\end{equation}
Hence, finite velocities affect the mesoscopic contact angle, and can consequently also lead to finite microscopic tip angles $\phi_0$ according to Eq.~(\ref{reno}).
  
\section{Determination of the eigenvalue}
\label{eigenval::appendix}

Here we give more details on the calculation of the eigenvalue $\mu$ in the regime $p \ll 1, \; \phi_0^2 \ll 1, \; p/\phi_0^2 \gg 1$ when $\beta/\mu \gg 1$. 
Specifically, the Laplace transformation of Eq.~(\ref{linEq})
 yields for the image space function of the slope 
\bea
\mathcal{L} \left[ \frac{dx}{d\xi} \right](s) = \frac{\mu s - 1 - \alpha s/(s + \beta \mu^{-1})}{\mu s^2 - \sqrt{\frac{2}{\pi}} s^{\frac{1}{2}}} =:f(s).
\eea
This function has singularities at $s^{\star}_1 = -\beta \mu^{-1}, s^{\star}_2 = (2/\pi)^{1/3} \mu^{-2/3}$ and $s^{\star}_3 = 0$. 
In general, the real space function $f(x)$ corresponding to an image space function $f(s)$ which exhibits a pole on the positive real axis at $s = a$ has a leading term $f(x)\sim \exp(a x)$. Consequently, the pole on the real positive axis at $s^{\star}_2$ is prohibited by the parabolic asymptotics of our interface, and the selection of the eigenvalue demands the compensation of that pole by a vanishing value of the numerator at $s=s^{\star}_2$. From the resulting quadratic equation in $\mu^{1/3}$, we pick the solution that has a proper behaviour as $\alpha \to 0$. Specifically, we demand that as $\alpha \to 0$, $\mu^{1/3} \to (\pi/2)^{1/3}$, which we know from our earlier considerations in \cite{EAB_CH_DP_DET_PRL99_105701_2007}.
From the obtained expression of the eigenvalue (\ref{mu_cubeRoot})
the explicit representation of the melting velocity in terms of the material parameters reads finally
\bea 
\frac{v d}{D} &=& \mu \frac{2}{\pi} \frac{\Delta^3}{\phi_0^3} \nonumber \\
&\sim& \Bigg[  \frac{\Delta+\Delta_{w}-2\frac{d}{\delta}\phi_0^2}{\phi_0} + 
\Bigg( \frac{\Delta^2}{\phi_0^2} + \frac{2 \Delta_{w} + 4 \frac{d}{\delta}\phi_0^2}{\phi_0} \Delta  \nonumber \\
&&+ \Bigg( \frac{\Delta_{w}}{\phi_0} - 2\frac{d}{\delta}\phi_0 \Bigg)^2 \Bigg)^{1/2} \Bigg]^3. \nonumber
\eea

\section{Determination of the interface}
\label{contour::app}

\begin{figure}
\begin{center}
\includegraphics[trim=4cm 4cm 9cm 3cm, clip=true,width=4.5cm]{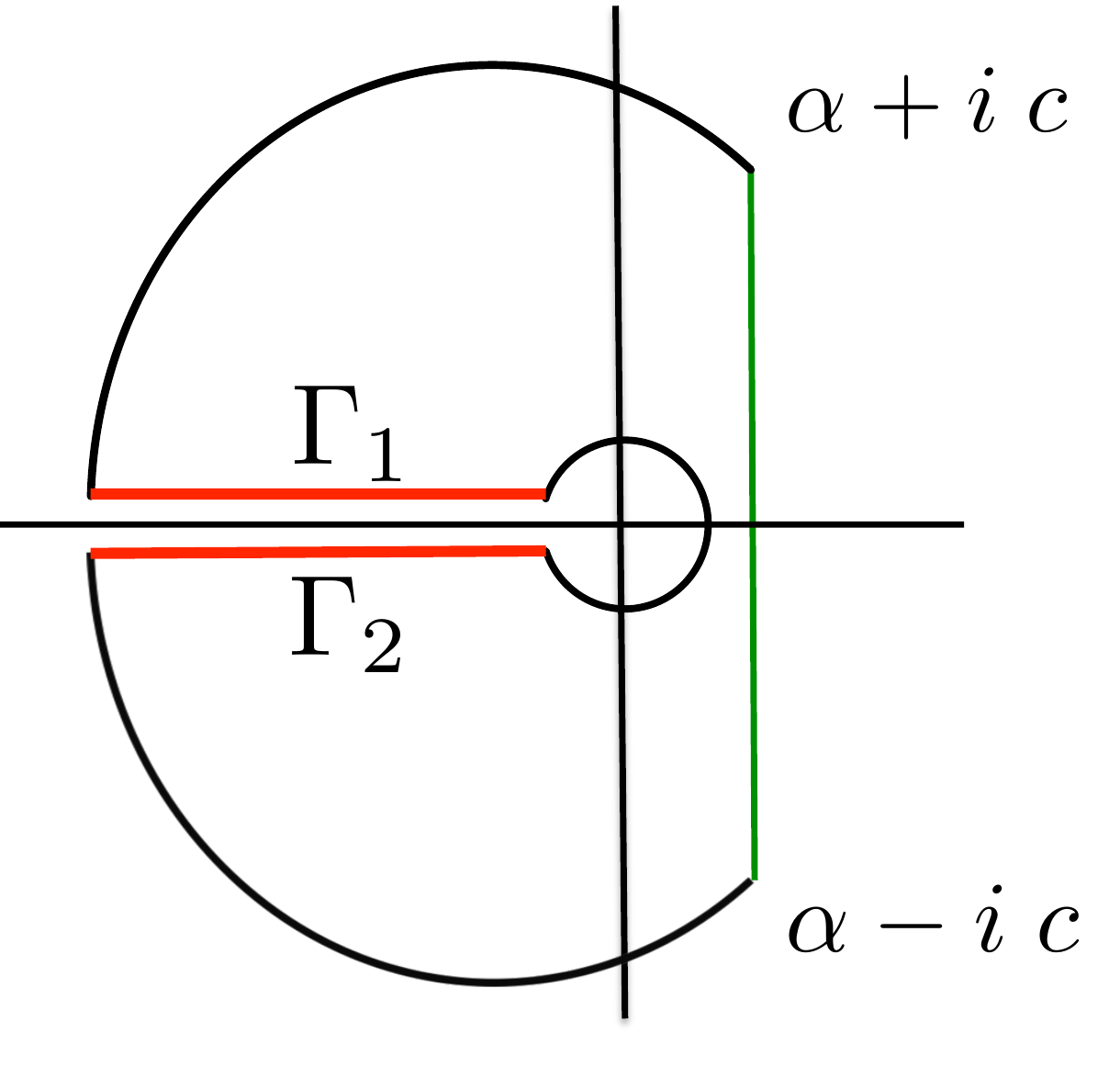}
\caption{\label{contour} 
The integration contour for evaluation of Eq.~(\ref{bromwichIntegral}). The evaluated integral is from $\alpha - ic$ to $\alpha + i c$, and the only contributions which contribute additionally for integration along the closed contour are $\Gamma_1$ and $\Gamma_2$, parallel to the negative real axis. In the limit $c \to \infty$, we thus can determine the inverse Laplace transform.}
\end{center}
\end{figure}

First, we describe an intermediate step for the analytic calculation of the slope of the interface. 
The inverse transformation of the Laplace transform of $dx/d\xi$  is obtained by the Bromwich integration, which reads 
\bea
\label{bromwichIntegral}
\frac{dx}{d\xi} = \frac{1}{2 \pi i} \int_{a - i \; \infty}^{a+i \; \infty} ds\, e^{s \xi} f(s).
\eea
Here, the real constant $a$ is chosen such that all singularities are located to the left of the vertical integration path at $\Re(s)=a$.
In our case, due to the selection mechanism, the function has remaining poles only on the 
negative real axis. We evaluate the integral via the residue theorem, applying one of the standard Bromwich contours --- specifically, we integrate the classical half circle, but exclude the branch cut on the negative real axis, as shown in Fig.~\ref{contour}.
We separate the Laplace transform $f(s)$ into contributions from the different singularities:
\bea
f(s) &=& \sqrt{\frac{\pi}{2}} \Bigg[ \frac{1}{\sqrt{s}} + \frac{c_2-c_3 \sqrt{s}-c_4 s}{s^{3/2}-c_5} \Bigg] + \frac{c_1}{\sqrt{\frac{\beta}{\mu}} +i \sqrt{s}} \nonumber \\
&&  + \frac{c_1^*}{\sqrt{\frac{\beta}{\mu}}-i \sqrt{s}}.  \label{newF}
\eea
Here we define
\bea
\tilde{\alpha} &=& \frac{\alpha}{\frac{2\mu}{\pi}+\beta^3}, \qquad c_1 = \frac{\tilde{\alpha}}{2} \left[ \sqrt{\beta^3 \mu} + i \mu \sqrt{\frac{2}{\pi}} \right], \nonumber \\ \nonumber
c_2 &=& \tilde{\alpha} \beta \frac{2}{\pi},\qquad c_3 = \left( \tilde{\alpha} - \beta^2 -1 \right) \sqrt{\frac{2}{\pi}}, \\ \nonumber
c_4 &=& \tilde{\alpha} \mu \frac{2}{\pi} + 1,\qquad c_5 = \frac{\sqrt{\frac{2}{\pi}}}{\mu}.
\eea
For the representation in Eq.~(\ref{newF}) we can directly evaluate several contributions,
\bea
\label{fromTables}
\frac{1}{\sqrt{s}} &\to& \frac{1}{\sqrt{\pi \xi}}, \nonumber  \\
\frac{1}{\sqrt{\frac{\beta}{\mu}}+i\sqrt{s}} &\to& -\frac{i}{\sqrt{\pi \xi}} + \sqrt{\frac{\beta}{\mu}} e^{-\frac{\beta}{\mu}\xi} \erfc\left[ -i \sqrt{\frac{\beta}{\mu}\xi} \right]. \nonumber
\eea
So from 
\begin{widetext}
\bea
\frac{dx}{d\xi} &=& \frac{1-i c_1 \sqrt{\frac{2}{\pi}}}{\sqrt{\frac{2}{\pi}} \sqrt{\pi \xi}} + \sqrt{\frac{\beta}{\mu}} c_1 e^{-\frac{\beta}{\mu}\xi} \erfc[-i\sqrt{\frac{\beta}{\mu}\xi}]  
+ c_1^* \left(\frac{i}{\sqrt{\pi \xi}} +\sqrt{\frac{\beta}{\mu}}  e^{-\frac{\beta}{\mu}\xi} \erfc[i\sqrt{\frac{\beta}{\mu}\xi}] \right) 
+ L^{-1} \left[ \frac{1}{\sqrt{\frac{2}{\pi}}} \frac{c_2-c_3 \sqrt{s} - c_4 s}{s^{3/2}-c_5} \right] \nonumber 
\eea
the only remaining integration is 
\bea
\label{realInt}
L^{-1} \left[ \frac{1}{\sqrt{\frac{2}{\pi}}} \frac{c_2-c_3 \sqrt{s} - c_4 s}{s^{3/2}-c_5} \right] 
&& = \lim_{\epsilon\to 0,\; R \to \infty} \frac{1}{i\sqrt{8 \pi}} \int_\epsilon^R \Bigg[ \frac{c_2 + i c_3 \sqrt{s}+c_4 s}{i s^{3/2}-c_5}
 - \frac{c_2 - i c_3 \sqrt{s}+c_4 s}{-i s^{3/2}-c_5} e^{-s \xi} \Bigg]\,ds \nonumber \\
 &&= \lim_{\epsilon\to 0,\; R \to \infty} -\sqrt{\frac{2}{\pi}}\int_\epsilon^R  \frac{c_4 q^6 + c_2 q^4 + c_5 c_3 q^2}{q^6+c_5^2}e^{-q^2 \xi} dq. \nonumber
\eea
After tedious algebraic manipulations we get the complete solution for $dx/d\xi$
\bea
\frac{dx}{d\xi} &=& -\frac{1}{3 \sqrt{2 \pi} c_5^2} \Bigg[ \\ \nonumber
&& \frac{c_2 \left( \pi c_5^{\frac{5}{3}} \left( \exp \left({\frac{3}{2} c_5^{\frac{2}{3}} \xi}\right) + \sqrt{3} \sin \left( \frac{1}{2} \sqrt{3} c_5^{\frac{2}{3}} \xi \right) + \cos \left(\frac{1}{2} \sqrt{3} c_5^{\frac{2}{3}} \xi \right) \right) - 6 \sqrt{\pi} c_5^{2} \sqrt{\xi} \exp\left( {\frac{1}{2} c_5^{\frac{2}{3}}} \xi \right) {}_1F_{3}\left( 1; \frac{1}{2}, \frac{5}{6}, \frac{7}{6}; \frac{c_5^2 \xi^3}{27}  \right) \right)}{\exp \left( \frac{1}{2} c_5^{\frac{2}{3}} \xi  \right)} \\ \nonumber
&+& c_5 c_3 \left( 4 \sqrt{\pi} c_5^2 \xi^{\frac{3}{2}} {}_1F_{3}\left( 1; \frac{5}{6}, \frac{7}{6}, \frac{3}{2}; \frac{c_5^2 \xi^3}{27}  \right) - \pi c_5 \exp \left( c_5^{\frac{2}{3}} \xi \right) + \frac{2 \pi c_5 \cos\left( \frac{1}{2} \sqrt{3} c_5^{\frac{2}{3}} \xi \right)}{\exp \left( \frac{1}{2} c_5^{\frac{2}{3}} \xi \right)} \right) \\ \nonumber
&-& \frac{c_5^2 c_4 \left( 5 \pi c_5^{\frac{1}{3}} \left( \exp\left( \frac{3}{2} c_5^{\frac{2}{3}} \xi \right) - \sqrt{3} \sin \left( \frac{1}{2} \sqrt{3} c_5^{\frac{2}{3}} \xi \right)  + \cos \left( \frac{1}{2} \sqrt{3} c_5^{\frac{2}{3}} \xi  \right)  \right) - 8 \sqrt{\pi} c_5^2 \xi^{\frac{5}{2}} \exp \left( \frac{1}{2} c_5^{\frac{2}{3}} \xi \right)  {}_1F_{3}\left( 1; \frac{7}{6}, \frac{3}{2}, \frac{11}{6}; \frac{c_5^2 \xi^3}{27}  \right) \right)}{\exp \left( \frac{1}{2} c_5^{\frac{2}{3}} \xi  \right)} \Bigg] \\ \nonumber
&+& c_1^\star \left( \frac{\sqrt{\frac{\beta}{\mu}} \erfc \left( i \sqrt{\xi} \sqrt{\frac{\beta}{\mu}}  \right) }{\exp \left( \xi \frac{\beta}{\mu} \right)} + \frac{i}{\sqrt{\pi} \sqrt{\xi}} \right) 
+ \frac{c_1 \sqrt{\frac{\beta}{\mu}}\erfc \left( -i \sqrt{\xi} \sqrt{\frac{\beta}{\mu}}\right)}{\exp \left( \xi \frac{\beta}{\mu} \right)} + \frac{1- i c_1 \sqrt{\frac{2}{\pi}}}{\sqrt{\pi}\sqrt{\frac{2}{\pi}} \sqrt{\xi}} - \frac{c_4}{\sqrt{2 \xi}}.
\eea
\end{widetext}
Here ${}_{i}F_{j}$ denotes the generalised hypergeometric function. 
This expression is then integrated numerically to obtain the solid-melt interface contours as shown in Fig.~\ref{geometry} and in excellent agreement with the direct numerical solution of Eq.~(\ref{BIRepresentation}).



\begin{thebibliography}{99}

\bibitem{MRappazAJacotWJBoettinger_MetallMaterTransA_34_467_2003} M. Rappaz, A. Jacot and W. J. Boettinger, Metall. Mater. Trans. A {\bf{34}} 467 (2003).

\bibitem{WFanXGong_PRB72_064121_2005} W. Fan and X. G. Gong, Phys. Rev. B {\bf{72}}, 064121 (2005).


\bibitem{AAdland_etal_PRB87_024110_2013} A. Adland, A. Karma, R. Spatschek, D. Buta and M. Asta, Phys. Rev. B {\bf{87}}, 024110 (2013).

\bibitem{RSpatschek_etal_PRB87_024109_2013} R. Spatschek, A. Adland and A. Karma, Phys. Rev. B {\bf{87}}, 024109 (2013).

\bibitem{NWang_etal_PRE81_051601_2010} N. Wang, R. Spatschek and A. Karma, Phys. Rev. E {\bf{81}}, 051601 (2010).

\bibitem{YMishin_etal_ActaMat57_3771_2009} Y. Mishin, W. J. Boettinger, J. A. Warren and G. B. McFadden, Acta Mat. {\bf{57}}, 3771 (2009).

\bibitem{TFrolovYMishin_PRL106_155702_2011} T. Frolov and Y. Mishin, Phys. Rev. Lett. {\bf{106}}, 155702 (2011).

\bibitem{EAB_CH_DP_DET_PRL99_105701_2007} E. A. Brener, C. H\"uter, D. Pilipenko, and D. E. Temkin, Phys. Rev. Lett {\bf{99}}, 105701 (2007).

\bibitem{Mullins}
W.W. Mullins, J. Appl. Phys. {\bf 28}, 333 (1957).

\bibitem{SnoijerAndreottiPhysFluid20_057101_2008} J. H. Snoijer and B. Andreotti, Phys. Fluids {\bf{20}}, 057101 (2008).

\bibitem{Fensin_2010_PRE_81_031601} S. J. Fensin et al, Phys. Rev. E {\bf{81}}, 031601 (2010).

\bibitem{LipowskiFisher_PRB36_2126_1987} R. Lipowsky and M. E. Fisher, Phys. Rev. B {\bf{36}}, 2126 (1987).

\bibitem{KikuchiCahn1980} R. Kikuchi and J. W. Cahn, Phys. Rev. B {\bf{21}}, 1893 (1980).

\bibitem{JMellenthinAKarmaMPlapp_PRB_78_184110_2008} J. Mellenthin, A. Karma, and M. Plapp, Phys. Rev. B {\bf{78}}, 184110 (2008).

\bibitem{deGennesRMP1985} P. G. de Gennes, Rev. Mod. Phys. {\bf{57}}, 827 (1985).

\bibitem{CHueter_GBoussinot_EABrener_PRE_83_050601_2011} C. H\"uter, G. Boussinot, E. A. Brener, and D. E. Temkin, Phys. Rev. E {\bf{83}}, 050601 (2011).

\bibitem{TFischaleckKKassner_EPL_81_54004_2008} T. Fischaleck and K. Kassner, Euro. Phys. Lett. {\bf{81}}, 54004 (2008).

\bibitem{JSLangerLATurski_ActaMetall_25_1113_1977} J. S. Langer and L. A. Turski, Acta Metall. {\bf{25}}, 1113 (1977).  

\bibitem{Ivantsov}
G. P. Ivantsov, Dokl. Akad. Nauk USSR {\bf {58}} (1947).

\bibitem{RNWenzel_IndEngChem28_p988_1936} R. N. Wenzel, Ind. Eng. Chem. {\bf{28}}, 988 (1936).

\bibitem{LGaoTJMcCarthy_Langmuir2007_23_3762} L. Gao and T. J. McCarthy, Langmuir {\bf{23}}, 3762 (2007).

\bibitem{AMarmur_AdvCollIntSci50_121} A. Murmur, Adv. Coll. Int. Sci. {\bf{50}}, 121 (1994).

\bibitem{GWolonskyAMarmurCollSurfA_156_p381_1999} G. Wolonsky and A. Marmur, Colloid and Surfaces A {\bf{156}}, 381 (1999).

\bibitem{Spatschek2010}
R. Spatschek and A. Karma, Phys. Rev. B. {\bf 81}, 214201 (2010).

\bibitem{EfimAIP}
E. A. Brener and V. I. Melnikov, Advances in Physics {\bf{40}} 53 (1991).



\end{thebibliography}
\end{document}